\documentclass[fleqn,usenatbib]{mnras}

\usepackage{newtxtext,newtxmath}

\usepackage[T1]{fontenc}

\DeclareRobustCommand{\VAN}[3]{#2}
\let\VANthebibliography\thebibliography
\def\thebibliography{\DeclareRobustCommand{\VAN}[3]{##3}\VANthebibliography}

\usepackage{graphicx}	
\usepackage{amsmath}	

\title[Hyperfine collisional excitation of NH$_3$]{Hyperfine collisional excitation of ammonia by molecular hydrogen}

\author[J. Loreau et al.]{
J. Loreau \thanks{E-mail: jerome.loreau@kuleuven.be}$^{1}$, A. Faure$^{2}$, F. Lique$^{3}$, S. Demes$^{3}$, P.J. Dagdigian$^{4}$
\\
$^1$ KU Leuven, Department of Chemistry, 3001 Heverlee, Belgium	\\
$^{2}$Univ. Grenoble Alpes / CNRS, Institut de Plan\'etologie et d'Astrophysique de Grenoble (IPAG) UMR 5274,  Grenoble F-38041, France		\\
$^3$ Univ Rennes, CNRS, IPR (Institut de Physique de Rennes) - UMR 6251, F-35000 Rennes, France \\
$^{4}$Department of Chemistry, The Johns Hopkins University, Baltimore, MD 21218-2685, USA.\\
}

\date{Accepted XXX. Received YYY; in original form ZZZ}

\pubyear{2023}

\begin{document}
\label{firstpage}
\pagerange{\pageref{firstpage}--\pageref{lastpage}}
\maketitle

\begin{abstract}
Ammonia is one of the most widely observed molecules in space, and many observations are able to resolve the hyperfine structure due to the electric quadrupole moment of the nitrogen nucleus. The observed spectra often display anomalies in the satellite components of the lines, which indicate substantial deviations from the local thermodynamic equilibrium. The interpretation of the spectra thus requires the knowledge of the rate coefficients for the hyperfine excitation of NH$_3$ induced by collisions with H$_2$ molecules, the dominant collider in the cold interstellar medium.
In this paper we present the first such calculations using a recoupling approach. The rate coefficients are obtained for all hyperfine levels within rotation-inversion levels up to $j=4$ and temperatures up to 100 K by means of quantum scattering close-coupling calculations on an accurate, five-dimensional, potential energy surface. 
We show that the rate coefficients depart significantly from those obtained with the statistical approach and that they do not conform to any simple propensity rules.
Finally, we perform radiative transfer calculations to illustrate the impact of our new rate coefficients by modelling the hyperfine line intensities of the inversion transition in ground state \textit{para}-NH$_3$ ($j_k=1_1$) and of the rotational transition $1_0\rightarrow 0_0$ in \textit{ortho}-NH$_3$.

\end{abstract}




\section{Introduction}

Since its first detection in the interstellar medium \citep{Cheung1968}, ammonia (NH$_3$) has often been used to probe the physical properties of molecular clouds due to the favourable structure of its energy levels and its high abundance in a variety of astrophysical sources. 
The rotational levels of NH$_3$ are denoted as $j_k$ (where $j$ is the angular momentum and $k$ its projection on the molecular axis), and the presence of metastable states for $j=k$ that can only decay through molecular collisions makes NH$_3$ an excellent probe of the local temperature. NH$_3$ exists in two forms, namely \textit{ortho}-NH$_3$ (for $k=3n$, $n$ integer) and \textit{para}-NH$_3$ (for $k=3n\pm 1$) that cannot be interconverted through radiative or inelastic collisional processes. In addition, the inversion (or umbrella) motion causes a splitting of each $j_k$ level into a doublet (with the exception of $k=0$ due to spin statistics) whose components, denoted as $j_k^+$ and $j_k^-$, are separated by 0.79 cm$^{-1}$ (23.68 GHz). NH$_3$ can thus be detected in different frequency ranges corresponding to pure inversion, rotational, or vibrational transitions.

In most astrophysical observations, the hyperfine structure due to the electric quadrupole moment of nitrogen is resolved. N nuclei possess a spin of $I=1$, so that each rotation-inversion level is further split into three hyperfine components $F$ (with the exception of ground state \textit{ortho}-NH$_3$), with a separation of the order of a MHz. 
A rotational line is thus split into five lines: the main line ($\Delta F =0$), as well as two inner and two outer satellite lines characterized by $\Delta F=\pm 1$. The lines are further split by the spins of the H nuclei, but this interaction is much weaker and will be neglected here.

The relative intensity of hyperfine lines can be used to derive opacities if the various hyperfine components have the same excitation temperature and if the opacity is not too large \citep{Barrett1977, Ho1983}. However, NH$_3$ hyperfine-resolved observations commonly display anomalies in the spectra, where the intensities within a pair of satellite lines are not equal. This indicates different excitation temperatures, meaning that the satellite components do not obey the local thermodynamic equilibrium (LTE) assumption in the interstellar medium. Such anomalies were first observed by \cite{Matsakis1977} in source DR21 and subsequently by various authors (see \textit{e.g.}, \cite{Matsakis1980, Gaume1996, Rathborne_2008, Rosolowsky_2008, Camarata2015, Caselli2017, Zhou2020}) in diverse sources.
If the hyperfine levels are not populated thermally, the interpretation of observations requires radiative transfer models to account for non-LTE effects. These occur mainly because of collisional excitation by H$_2$ molecules, and a knowledge of accurate rate coefficients for the hyperfine excitation of NH$_3$ by H$_2$ is thus necessary. In addition, modelling the strength of the hyperfine anomalies can be used to deduce the local H$_2$ density.

Models to interpret anomalies were developed by \cite{Stutzki1985a,Stutzki1985b} based on rate coefficients calculated for NH$_3$-He collisions as a proxy for NH$_3$-H$_2$ collisions. Beside this work, \cite{Chen1998} also computed hyperfine-resolved rate coefficients for NH$_3$-He collisions. To this day there are thus no accurate colisional data for the hyperfine excitation of NH$_3$ by H$_2$ molecules.

Various theoretical techniques exist to compute such rate coefficients. Among these, the simplest is the statistical approach, which assumes that the rate coefficients for hyperfine excitation are proportional to the hyperfine degeneracy of the final state. In this case, rate coefficients for pure rotational transitions, which are available for molecules such as NH$_3$, are the only ingredient needed to obtain hyperfine-resolved rate coefficients. This method was used to interpret the observations of \cite{Caselli2017} for the $1_0-0_0$ transition in the pre-stellar core L1544. The infinite-order-sudden limit approximation, used by \cite{Stutzki1985a}, ignores the rotational energy spacing and allows again to obtain the hyperfine-resolved rate coefficients on the basis of the pure rotational ones. Given the nature of its assumptions, this approximation is however not expected to be valid at low temperatures, i.e. at temperatures close or lower than the rotational spacings. Finally, the method used in the present work is the recoupling approach, which was also adopted by \cite{Chen1998} but for NH$_3$-He collisions and combined with the coupled-states approximation\footnote{We note that the results of \cite{Chen1998} are questionable and should not be used because some of their rate coefficients are unphysically large (up to $10^{-7}$ cm$^3$s$^{-1}$).}. The only approximation in our version of the recoupling approach is the assumption that the hyperfine levels are degenerate, i.e. their splitting is negligible compared to the temperature \citep{Alexander1985}. While computationally more demanding, it is also expected to be more accurate than the statistical or sudden methods. A summary and comparison of these different approaches can be found in the paper by \cite{Faure2012} for the linear molecules CN and HCN.

In this work we present rate coefficients for the hyperfine excitation of \textit{ortho}- and \textit{para}-NH$_3$ in collisions with \textit{para}-H$_2$ for all levels up to $j=4$ and for temperatures up to 100 K. While the rotation-inversion collisional excitation of NH$_3$ by H$_2$ has been widely studied theoretically over the past decades (\cite{Danby1987, Offer1990, Maret2009, Bouhafs2017,  Ma2015a, Demes2023a}) and experimentally with crossed beams \citep{Schleipen1991, Tkac2015b, Gao2019a}, to the best of our knowledge this represents the first effort to provide hyperfine collisional excitation rate coefficients of NH$_3$ (or any non-linear molecule) by H$_2$ with the accurate recoupling approach.

The paper is structured as follows: in Section \ref{sec_theory} we discuss the theoretical methods employed. In Section \ref{sec_rate_coeffs} we discuss the features of the hyperfine-resolved rate coefficients, and in Section \ref{sec_rad_trans} we present radiative transfer calculations to investigate the impact of these new data on the modelling of ammonia spectra. Finally, we summarize our findings in Section \ref{sec_concl}.


\section{Theory}\label{sec_theory}

The scattering calculations were performed in two steps. First, we computed the scattering matrix for rotation-inversion transitions by means of the fully quantum-mechanical close-coupling (CC) method. This method has been used to obtain accurate cross sections for rotational excitation of NH$_3$ induced by collisions with H$_2$ by several authors (see \textit{e.g.}, \cite{Danby1987,Rist1993}) and we refer to these papers for further methodological details. Similarly to recent theoretical works \citep{Ma2015a,Bouhafs2017,Demes2023a}, we employ the accurate five-dimensional rigid-rotor potential energy surface of \cite{Maret2009}, which was constructed by fitting \textit{ab initio} energies obtained with the coupled cluster method with single, double, and perturbative triple excitations.

The cross sections for rotation-inversion transitions can be obtained from the $T$ matrix elements $T^J_{j k v j_2 j_{12} l , j' k' v' j'_2 j'_{12} l' }$, where $v$ denoted the inversion level, $J$ is the quantum number associated with the total angular momentum of the collision complex without nuclear spin ($\boldsymbol{J}=\boldsymbol{j_{12}}+\boldsymbol{L}$), $\boldsymbol{j_{12}}=\boldsymbol{j_1}+\boldsymbol{j_2}$, where $\boldsymbol{j_1}$ and $\boldsymbol{j_2}$ are the angular momenta of NH$_3$ and H$_2$, respectively, and $\boldsymbol{L}$ is the relative orbital angular momentum of the collision. We do not discuss here the properties of the cross sections for rotation-inversion transitions as these are identical to those reported recently by \cite{Bouhafs2017} and \cite{Demes2023a}.

In a second step, the recoupling approach is used to compute hyperfine-resolved cross sections. When the nuclear spin of the N atom ($I=1$) is included, each level of NH$_3$ is split into three components, and the total angular momentum $\boldsymbol{J_t}$ is obtained by the coupling $\boldsymbol{J_t}=\boldsymbol{J}+\boldsymbol{I}$. However, the hyperfine splittings are of the order of the MHz (i.e. about $10^{-5}-10^{-4}$ K), which is much smaller than the rotational spacings and collisional energies considered here. We thus assume that the hyperfine levels of NH$_3$ are degenerate. In this case the nuclear spin wavefunctions can be decoupled from the rotational wavefunctions and, as shown by \cite{Offer1994}, the angular momenta can also be recoupled as $\boldsymbol{F}=\boldsymbol{j_1}+\boldsymbol{I}$, $\boldsymbol{j_R}=\boldsymbol{j_2}+\boldsymbol{L}$, $\boldsymbol{J_t}=\boldsymbol{j_R}+\boldsymbol{F}$. The $T-$matrix elements including nuclear spin, $T^{J_t}$, can then be obtained from the spin-free T-matrix elements $T^J$ as
\begin{multline}
T^{J_t}_{j v F j_2 j_R l , j' v' F' j'_2 j'_R l'}=\sum_{J, j_{12}, j'_{12}} (-1)^{j_R+j'_R+l+l'+j_2+j'_2} 
 \\
 ([F][F'][j_{12}][j'_{12}][j_R][j'_R])^{1/2}[J]  
\left\{\begin{array}{ccc}
 j & j_2 & j_{12} \\
 l & J & j_R 
\end{array}\right\}
\left\{\begin{array}{ccc}
j' & j'_2 & j'_{12} \\
 l' & J & j'_R 
\end{array}\right\}		 \\
\left\{\begin{array}{ccc}
 j_R & j & J \\
 I & J_t & F 
\end{array}\right\}
\left\{\begin{array}{ccc}
 j'_R & j' & J \\
 I & J_t & F' 
\end{array}\right\}
T^J_{j k v j_2 j_{12} l , j' k' v' j'_2 j'_{12} l' }
\end{multline}
where $\{\cdots\}$ denotes a 6-j symbol and $[x]=2x+1$.

The state-to-state hyperfine cross sections for NH$_3$-H$_2$ can be then computed from the $T$-matrix according to the expression \citep{Offer1994}:
\begin{multline} \label{ICS_eq}
 \sigma_{j k v F j_2 , j' k' v' F' j_2'}=\frac{\pi}{k_i^2(2F+1)(2j_2+1)}\sum_{J_{T}}(2J_{T}+1) \\
 \sum_{l l' j_R j'_R} \vert T^{J_t}_{jk v F j_2 j_R l , j' k' v' F' j'_2 j'_R l}\vert^2 
\end{multline}

The scattering calculations were carried out with the \texttt{Hibridon} program \citep{Hibridon2023} for total energies between 0 and 600 cm$^{-1}$ on a grid of 700 energies for \textit{ortho}-NH$_3$ and 660 energies for \textit{para}-NH$_3$. These calculations were performed independently for \textit{ortho}-NH$_3$ and \textit{para}-NH$_3$ as \textit{ortho}-para conversion is forbidden in inelastic collisions. 
The energy step was 0.5 cm$^{-1}$ at the lowest energies and increased gradually up to 4 cm$^{-1}$ at the highest energies. We used the log-derivative propagator in the short range, and the Airy propagator in the long range ($R>25 a_0$). Transitions between levels up to $j_1=6$ with $j_2=0$ were converged with a rotational basis including all levels up to $j_1=8$ and $j_2=0,2$. 
The rotational constants of NH$_3$ were taken as $A=B=9.9402$ cm$^{-1}$, $C=6.3044$ cm$^{-1}$ with an  inversion splitting of 0.7903 cm$^{-1}$, while the rotational constant of H$_2$ was taken as 59.3801 cm$^{-1}$ as in \cite{Bouhafs2017}. Tests were performed to ensure convergence of the cross sections with respect to $J_t$ in Eq. (\ref{ICS_eq}) at all energies, as well with respect to the parameters of the radial grid.

Rate coefficients were then calculated by averaging cross sections for initial and final states $i$ and $f$, $\sigma_{i f} (E_c$),  over the Maxwell-Boltzmann distribution of collision energies as
expressed :
\begin{equation}
 k_{i f}(T)=\biggl(\frac{8}{\pi\mu\beta}\biggl)^{\frac{1}{2}}\beta^2\int_0^{\infty} E_c \sigma_{i f}(E_c)e^{-\beta E_c} dE_c
\end{equation}
Where $\beta$=$\frac{1}{k_BT}$ and $k_B$, $T$ and $\mu$ denote the Boltzmann constant, the kinetic temperature and the NH$_3$-H$_2$ reduced mass, respectively.\\


\section{Hyperfine rate coefficients}\label{sec_rate_coeffs}

Hyperfine excitation rate coefficients were calculated for temperatures up to 100 K and for transitions involving all levels up to $j_1=4$ for both \textit{ortho}-NH$_3$ and \textit{para}-NH$_3$, corresponding to a total of 625 transitions for \textit{ortho}-NH$_3$ and 2304 transitions for \textit{para}-NH$_3$.
Only excitation due to p-H$_2$ ($j_2=0$) is considered here since excited H$_2$ levels are not significantly populated at the temperatures of molecular clouds where the hyperfine structure is resolved.

\begin{figure}
\includegraphics[width=1\linewidth]{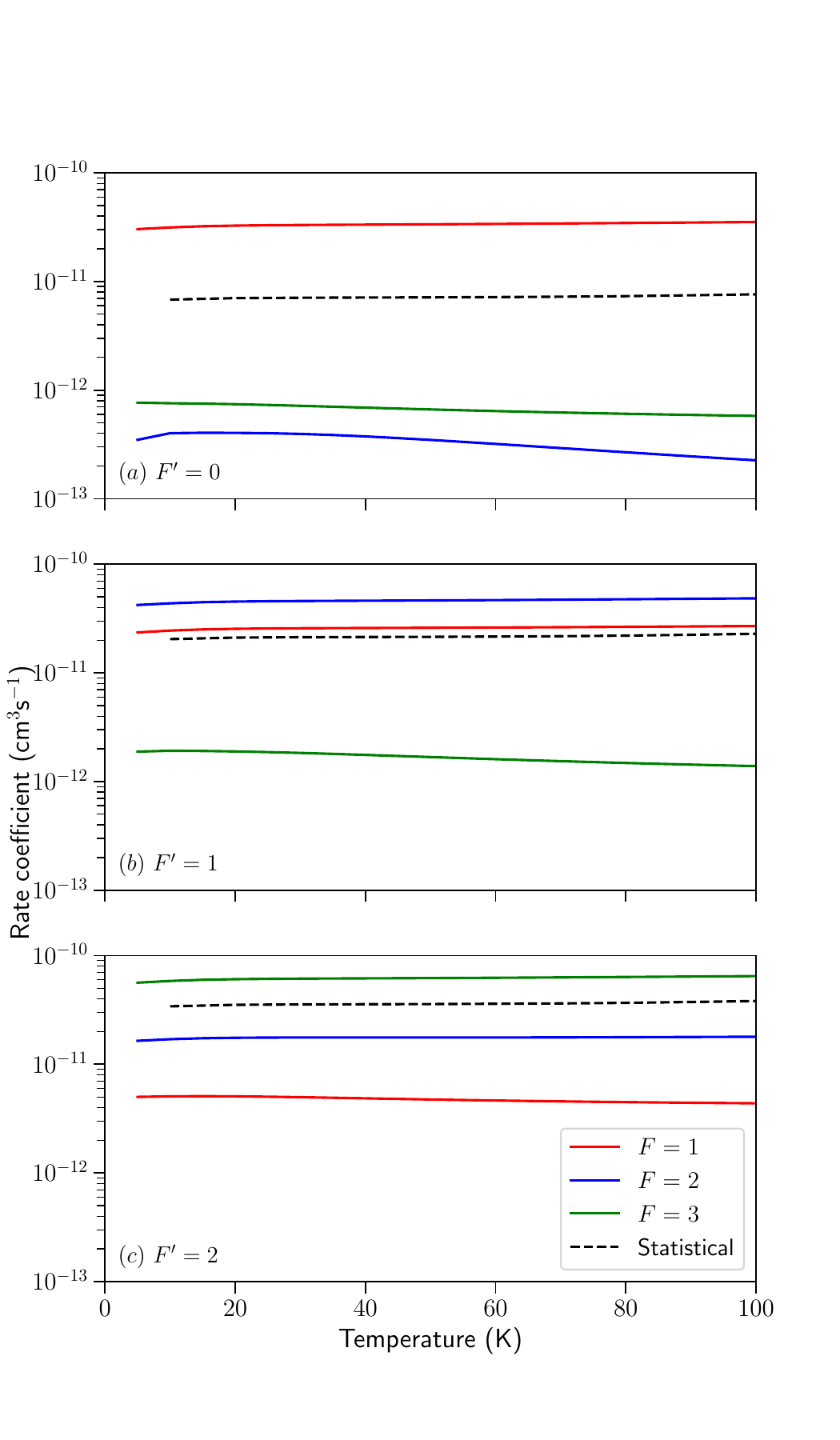}	\\
\caption{Rate coefficients for the transitions $2_1^+F \rightarrow 1_1^-F'$ of NH$_3$ induced by collisions with H$_2$ as a function of temperature. Panel (a), $F'=0$; panel (b), $F'=1$; panel (c), $F'=2$, and comparison to the statistical rate coefficients given by Eq. (\ref{hfs_stat}).}
\label{rates_functionT} 
\end{figure}

Rate coefficients for hyperfine transitions from the rotational state $2_1^+$ down to level $1_1^-$ are presented in Fig. \ref{rates_functionT} as a function of temperature up to 100 K. The rate coefficients are seen to be weakly dependent on $T$, a trend that follows directly from the behaviour of the rate coefficients for pure rotational transitions. 
On the same figure, we also present for comparison the statistical hyperfine rate coefficients. The statistical approach, also known as the $M_J$ randomizing limit \citep{Alexander1985}, is a commonly used approximation to compute hyperfine-resolved excitation rate coefficients, which assumes that the cross sections or rate coefficient for hyperfine excitation are directly proportional to the degeneracy of the final hyperfine state and independent of the initial hyperfine state. For NH$_3$-H$_2$ the statistical rate coefficients are expressed as
\begin{equation}\label{hfs_stat}
k_{jkvF, j'k'v'F'}^{\textrm{stat}}= \frac{2F'+1}{(2I+1)(2j'+1)}k_{jkv, j'k'v'}
\end{equation}
where $k_{jkv, j'k'v'}$ are the nuclear spin-free rate coefficients calculated with the CC method and $I=1$. To simplify the notation, in the remainder of the paper, we use $j$ to denote the rotational quantum number of NH$_3$ instead of $j_1$ without ambiguity since $j_2=0$.

As can be seen from Fig. \ref{rates_functionT}, for all three final hyperfine states of the transition $2_1^+F \rightarrow 1_1^-F'$ the statistical rate coefficients fall between the recoupling values, but a strong dependence on $F'$ can be observed and large deviations from the $M_J$ randomizing limit are seen to occur. It is thus clear that the statistical method cannot be used to obtain accurate hyperfine rate coefficients in the case of NH$_3$-H$_2$ collisions. 

From  Fig. \ref{rates_functionT}, it can be seen that the known propensity rule $\Delta F = \Delta j$ \citep{Alexander1985} holds for the transitions within $2_1^+ \rightarrow 1_1^-$. However, this is not a general trend. In Table \ref{table_21p} we report the rate coefficients for transitions from levels $2_1^+F$ down to levels $1_1^+F$ and $1_1^-F$ at a fixed temperature of 30 K. For the case of $2_1^+F$ to $1_1^+F$ transitions (as well as for a large number of transitions involving higher lying rotational levels), the $\Delta F = \Delta j$ propensity rule is not respected. In addition, Table \ref{table_21p} also shows that the largest rate coefficients are not necessarily those with the largest $F'$, as would be expected from a statistical behavior.

Of interest are also the quasi-elastic rate coefficients, i.e. rate coefficients for hyperfine excitation within a single rotational level. We present in Table \ref{table_quasi_el} the quasi-elastic rate coefficients $j_kvF \rightarrow j_kvF'$ at 25 K for the first three rotational states of \textit{para}-NH$_3$, and for the antisymmetric inversion level ($v=v'=-$). The corresponding $F-F'$ transitions involving the symmetric inversion levels ($v=v'=+$) are similar although not identical, with differences of up to a few percent (not shown). In addition, we also report in the same table the rate coefficients corresponding to a change of inversion level ($- \rightarrow +$; the reverse rate coefficients can be obtained by detailed balance). It should be noted that those are not strictly quasi-elastic, since the $- \rightarrow +$ pure inversion transition correspond to an energy transfer of 0.79 cm$^{-1}$. It can be noticed that the rate coefficients strongly depend on the final inversion level, with differences that reach a factor 5 between $v'=-$ and $v'=+$. Interestingly, the rate coefficients for inversion-conserving transitions $j_k^-F \rightarrow j_k^-F'$, which would be elastic if the nuclear spin were neglected, are in many cases smaller than those of the corresponding inversion-changing transitions $j_k^-F \rightarrow j_k^+F'$.
Finally, the last column of Table \ref{table_quasi_el} compares our values to the results of \cite{Stutzki1985a}, which are independent of the inversion level, and show important discrepancies of up to an order of magnitude for some quasi-elastic transitions. It is worth emphasizing that those results were obtained using a different method, namely the infinite order sudden (IOS) approximation, which is not expected to be accurate at low temperatures. In addition, \cite{Stutzki1985a} used a potential for NH$_3$-He instead of NH$_3$-H$_2$, which leads to an additional error. Finally, we note that our quasi-elastic rate coefficients are of the same order of magnitude as those for inelastic transitions.

\begin{table}
\centering
\caption{Rate coefficients for the transitions $2_1^+F \rightarrow 1_1^+F'$ and $2_1^+F \rightarrow 1_1^-F'$ of NH$_3$ induced by H$_2$ at a temperature of 30 K. Brackets indicate powers of 10.}\label{table_21p}
\begin{tabular}{ccccc}
\hline
\hline
$j_kv-j'_{k'}v'$ 		& 		& $F'=0$ 	& $F'=1$		& $F'=2$  \\ \hline
$2_1^+ - 1_1^+$	& $F=1$	& 5.16[-13]	& 1.98[-11]	& 7.27[-12]	\\
				& $F=2$	& 8.62[-12]	& 3.11[-12]	& 1.58[-11]		\\
				& $F=3$	& 1.85[-13]	& 8.99[-12]	& 1.84[-11]		\\	\hline
$2_1^+ - 1_1^-$	& $F=1$	& 3.31[-11]	& 2.57[-11]	& 4.97[-12]	\\
				& $F=2$	& 3.96[-13]	& 4.58[-11]	& 1.76[-11] 	\\
				& $F=3$	& 7.19[-13]	& 1.83[-12]	& 6.12[-11]	\\
\hline
\end{tabular}
\end{table}

\begin{table}
\centering
\caption{Quasi-elastic rate coefficients for the excitation $j_kvF \rightarrow j_kv'F'$ for rotational levels $1_1$, $2_2$, and $2_1$ of NH$_3$ by H$_2$ at 25 K. Comparison with the data of \citet{Stutzki1985a} (last column). Brackets indicate powers of 10.}\label{table_quasi_el}
\begin{tabular}{ccccc}
\hline
\hline
$j_k$ 		& $F-F'$ 	& $v=-, v'=-$ 	& $v=-, v'=+$	& Ref.  \\ \hline
$1_1$	& $0-2$	& 2.43[-11]	& 1.22[-11]	& $2.49[-11]$	\\
$1_1$	& $0-1$	& 1.83[-11]	& 8.96[-11]	& $8.66[-11]$	\\
$1_1$	& $2-1$	& 1.55[-11]	& 2.79[-11]	& $3.28[-11]$	\\
$2_2$	& $1-3$	& 2.58[-12]	& 1.42[-12]	& $1.92[-11]$	\\
$2_2$	& $1-2$	& 1.44[-11]	& 3.05[-11]	& $7.02[-11]$	\\
$2_2$	& $3-2$	& 7.42[-12]	& 1.41[-11]	& $3.88[-11]$	\\
$2_1$	& $2-3$	& 5.31[-12]	& 8.29[-12]	& $5.26[-11]$	\\
$2_1$	& $2-1$	& 4.13[-12]	& 6.84[-12]	& $4.14[-11]$	\\
$2_1$	& $3-1$	& 7.93[-13]	& 9.17[-13]	& $7.41[-12]$	\\
\hline
\end{tabular}
\end{table}

From their IOS calculations, \citet{Stutzki1985a} compute relative factors, defined as the ratio of hyperfine rate coefficients divided by the rotational excitation rate coefficient, 
\begin{equation}\label{eq_rel_factors}
g_{FF'}=\frac{k_{jkvF, j'k'v'F'}}{k_{jkv, j'k'v'}}
\end{equation}
While the absolute IOS rate coefficients are not expected to be accurate at low temperature, the method should provide relative factors that are better than the statistical ones. These relative factors obtained with the recoupling approach are presented in Table \ref{table_rel_factors} and compared to the IOS results of \citet{Stutzki1985a} for the $1_1^+ - 2_1^+$ and $1_1^+ - 2_1^-$ excitation transitions. We observe that in most cases the IOS method does predict the correct propensities, and performs better than the statistical approach, as expected. On the other hand, it is clear that IOS is not sufficient to obtain accurate rate coefficients, since discrepancies of up to 50\% are seen for the dominant rate coefficients and up to a factor 10 for the smallest rate coefficients. Moreover, both the IOS and recoupling relative factors are almost independent of the temperature in the range explored here (see Fig. \ref{rates_functionT}), which shows that the discrepancies discussed above remain up to temperatures of at least 100~K.

\begin{table}
\centering
\caption{Relative factors (Eq. (\ref{eq_rel_factors})) for the transitions $1_1^+F \rightarrow 2_1^+F'$ and $1_1^+F \rightarrow 2_1^-F'$ of NH$_3$ induced by H$_2$ at a temperature of 25 K. The values in parentheses are from \citet{Stutzki1985a}.}\label{table_rel_factors}
\begin{tabular}{ccccc}
\hline
\hline
				& 		& $F'=1$ 		& $F'=2$		& $F'=3$  \\ \hline
$1_1^+F - 2_1^+F'$	& $F=0$	& 0.03 (0.08)	& 0.95 (0.65)	& 0.02 (0.27)	\\
				& $F=1$	& 0.43 (0.31)	& 0.11 (0.20)	& 0.46 (0.48)		\\
				& $F=2$	& 0.09 (0.16)	& 0.35 (0.35)	& 0.56 (0.50)		\\	\hline
$1_1^+F - 2_1^-F'$	& $F=0$	& 0.93 (0.57)	& 0.02 (0.28)	& 0.05 (0.15)	\\
				& $F=1$	& 0.24 (0.27)	& 0.72 (0.50)	& 0.04 (0.23)	\\
				& $F=2$	& 0.03 (0.09)	& 0.17 (0.24)	& 0.81 (0.67)	\\
\hline
\end{tabular}
\end{table}


\section{Radiative transfer calculations}\label{sec_rad_trans}

In order to investigate the impact of the hyperfine rate coefficients
on the non-LTE modelling of NH$_3$ spectra, we present below simple
radiative transfer calculations using two different sets of
hyperfine-resolved collisional data: (1) the precise set of {\it
  recoupling} rate coefficients and (2) the approximate set of {\it
  statistical} rate coefficients, as defined above in Eq.~(\ref{hfs_stat}).

Radiative transfer calculations were performed with the non-LTE
\texttt{RADEX} program \citep{vandertak07} using the large velocity
gradient (LVG) formalism for an expanding sphere. The code was
employed to compute the radiation temperature $T_R$ of low-lying
hyperfine transitions in both \textit{para}- and \textit{ortho}-NH$_3$. The hyperfine
energy levels and radiative rates were taken from the CDMS catalog
\citep{muller05}.  The kinetic temperature was fixed at 10~K, which is
typical of cold molecular clouds, while the hydrogen density was
varied in the range $10^2-10^{9}$~cm$^{-3}$. The \textit{para}-NH$_3$ (or
\textit{ortho}-NH$_3$) column densities were selected in the range
$10^{12}-10^{15}$~cm$^{-2}$ with a line width (full width at half
maximum) of 1~km~s$^{-1}$. No radiation field was considered except
the cosmic microwave background (CMB) at 2.73~K. We have also checked
with the \texttt{MOLPOP} code \citep{asensio18} that line overlap
effects, which are ignored by \texttt{RADEX}, are mostly negligible
with the above parameters, allowing us to focus on the impact of the
collisional data.

We first consider the inversion transition $1_1^-\to 1_1^+$ of
\textit{para}-NH$_3$ at 23.68~GHz. This transition has been widely observed in
the interstellar medium since its first detection by
\cite{Cheung1968}. Anomalies in the hyperfine intensities of this line
were reported in the 1970s \citep{Matsakis1977} and these anomalies are
now known to be very common in star forming regions. In the anomalous
spectra, the intensities of the outer satellite lines are not equal:
 the $F=0\to 1$ component is stronger and the $F=1\to 0$ component
is weaker than expected. Radiative transfer models have shown that
these anomalies can be reproduced by non-LTE effects induced by photon
trapping of selected hyperfine transitions in the rotational
transition $2_1^{+/-}\to 1_1^{-/+}$, as explained by \cite{Stutzki1985b}
(see in particular their Fig.~2). Systematic motions, such as
expansion and collapse, are also expected to play a role and the
relative contribution of these effects remains a matter of debate
\citep[see][and references therein]{Zhou2020}. In any case, because
non-LTE populations depend on the competition between the radiative
and collisional rates, the accuracy of hyperfine selective collisional
rate coefficients should be critical for a correct interpretation of
the anomalies in the $1_1^-\to 1_1^+$ inversion transition.

\begin{figure}
\centering{\includegraphics[width=10cm]{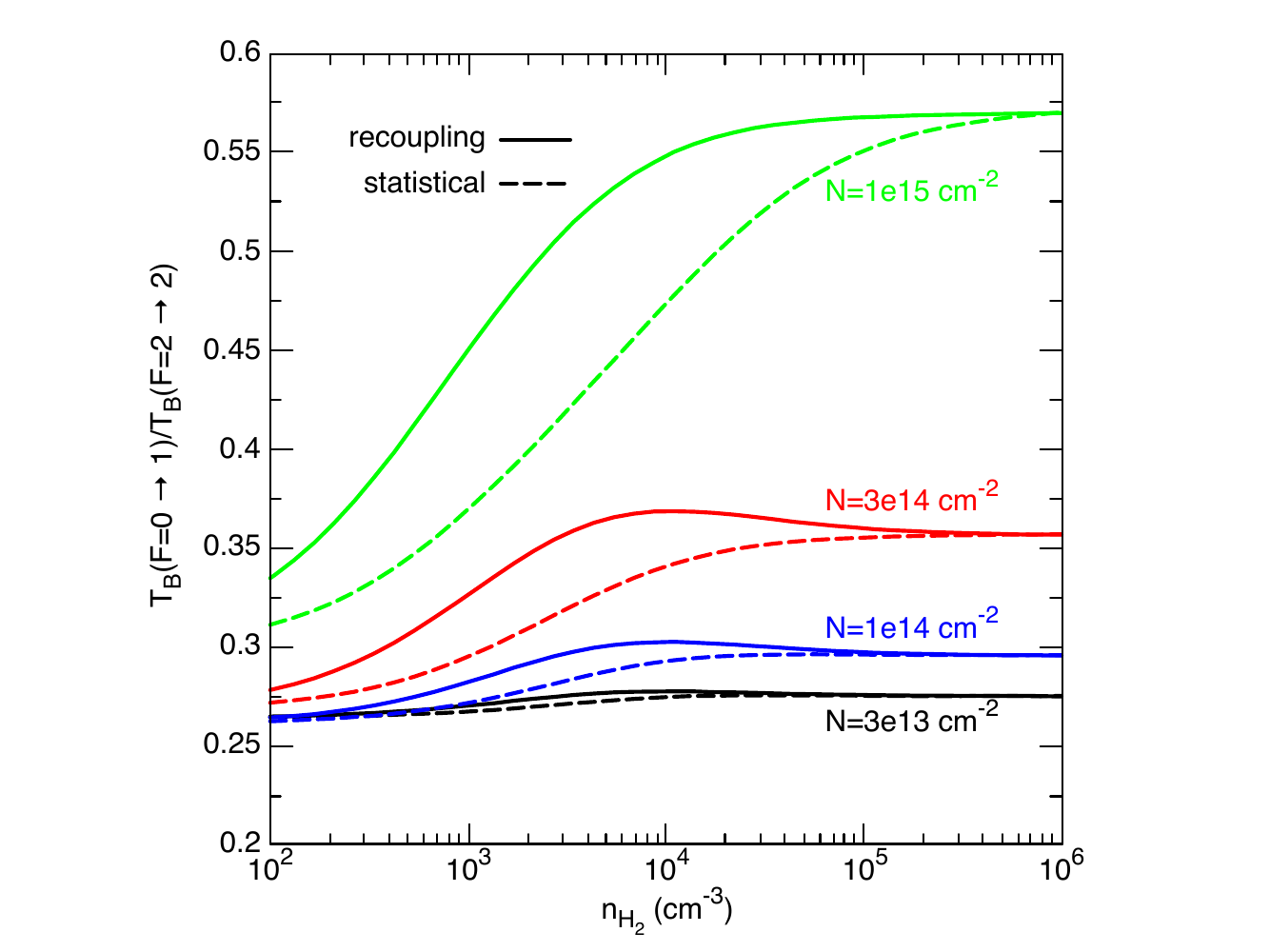}}
\caption{The intensity ratio of the hyperfine satellite line $F=0\to
  1$ of the \textit{para}-NH$_3$ $j_k=1_1$ inversion transition relative to the
  strongest component $F=2\to 2$ as function of the hydrogen density
  for a set of column densities. The kinetic temperature is fixed at
  10~K and the line width at 1~km~s$^{-1}$.}\label{fig:ratio_11}
\end{figure}

In Fig.~\ref{fig:ratio_11}, we have plotted the intensity ratio
of the outer hyperfine satellite line $F=0\to 1$ relative to the
strongest component $F=2\to 2$, for the two sets of collisional rates
(recoupling and statistical). At low hydrogen density, the ratio is
found to increase above the ``natural'' intensity ratio (0.266) as
soon as the column density exceeds $\sim 10^{14}$~cm$^{-2}$. The
ratios then increase with increasing density and plateaus are reached
when $n({\rm H_2})$ is larger than about $10^5$~cm$^{-3}$, where LTE
conditions are fullfilled. For column densities above $\sim
10^{14}$~cm$^{-2}$, we can notice the impact of the
collisional dataset: the intensity ratios are decreased by up to 20\%
when the statistical dataset is employed. This demonstrates the
importance of the hyperfine propensity rules when the total opacity of
the line is larger than $\sim 1$. A similar result was actually
observed by \cite{Stutzki1985a} using hyperfine collision rates computed
with the IOS approximation and using Helium
instead of H$_2$ as the collider. As concluded by these authors, a
precise knowledge of the hyperfine-resolved rate coefficients is thus
required for a reliable modelling of observational data.

\begin{figure}
\centering{\includegraphics[width=10cm]{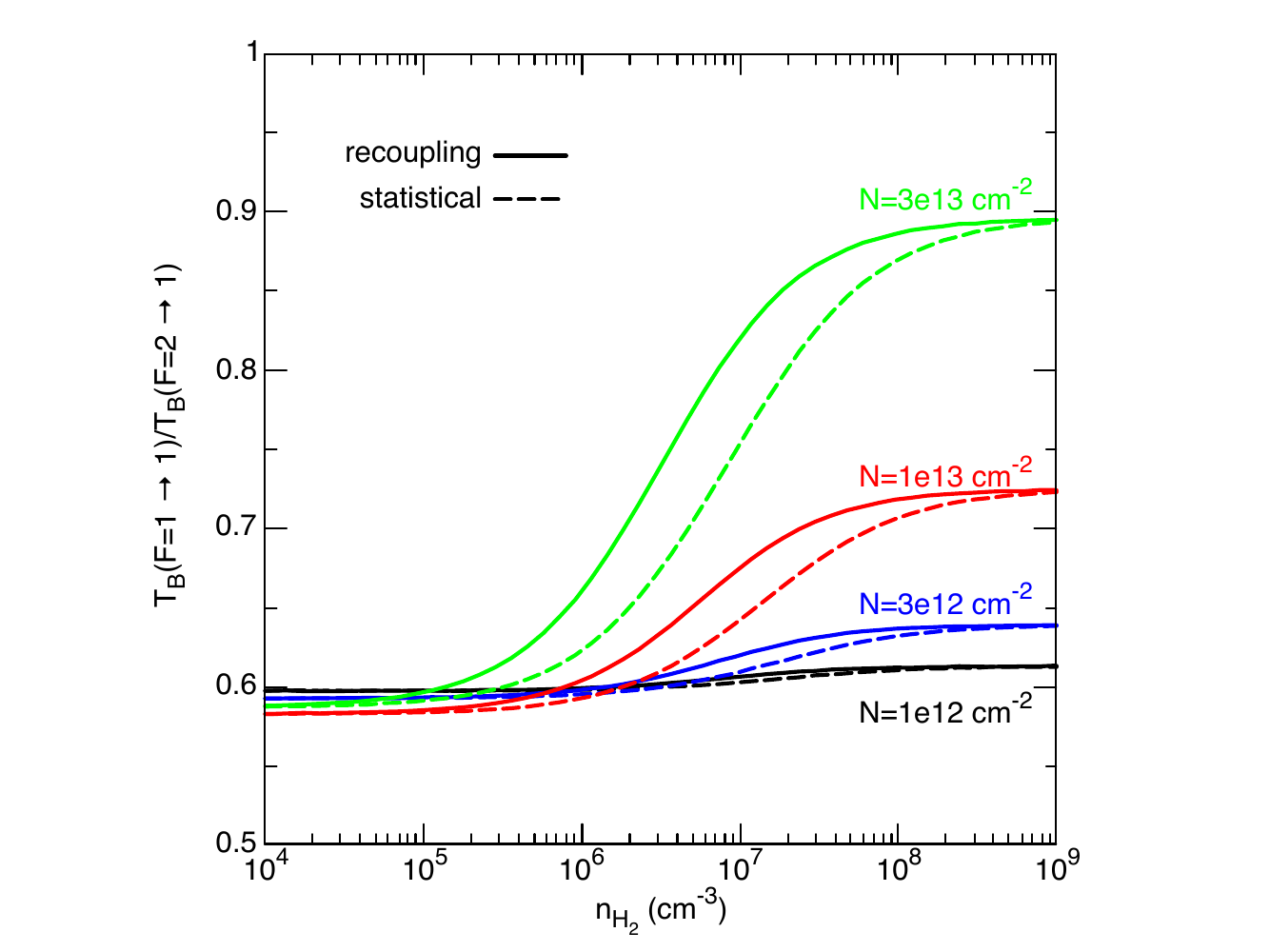}}
\caption{The intensity ratio of the hyperfine component $F=1\to 1$ of
  the \textit{ortho}-NH$_3$ $j_k=1_0\to 0_0$ rotational transition relative to
  the strongest component $F=2\to 1$ as function of the hydrogen
  density for a set of column densities. The kinetic temperature is
  fixed at 10~K and the line width at
  1~km~s$^{-1}$.}\label{fig:ratio_10}
\end{figure}

A second example is the \textit{ortho}-NH$_3$ ground-state rotational
transition $1_0^+ \to 0_0^-$ at 572.5~GHz. The electric quadrupole hyperfine
structure of this line was resolved for the first time in space thanks
to {\it Herschel} observations towards the cold pre-stellar core L1544
\citep{Caselli2017}. The three hyperfine components were found to be
heavily self-absorbed and a sophisticated radiative transfer model was
employed to reproduce the observed profile and to deduce the NH$_3$
radial abundance in the core. In that work, hyperfine selective
collisional data were generated from the rotational rate coefficients
of \citet{Bouhafs2017} by applying the statistical approach to the
electric quadrupole and magnetic hyperfine splittings. In
Fig.~\ref{fig:ratio_10}, we have plotted the intensity ratio of the
hyperfine component $F=1\to 1$ relative to the strongest component
$F=2\to 1$, again for the two sets of collisional rates (recoupling
and statistical). We note that our statistical set is identical to
that used by \cite{Caselli2017} except that the magnetic splittings
($<100$~kHz) are neglected here. At low hydrogen density, the ratio is
found to lie slightly below the natural intensity ratio (0.6) for
column density above $\sim 10^{12}$~cm$^{-2}$. When the H$_2$ density
exceeds $\sim 10^5$~cm$^{-3}$, however, the intensity ratios rise
above 0.6 and the LTE plateaus are reached at H$_2$ densities larger
than $\sim 10^8$~cm$^{-3}$, as expected from the large radiative rates
($A\sim 1.58\times 10^{-3}$). The impact of the rate coefficients is
smaller than in the case of the \textit{para}-NH$_3$ transition
(Fig.~\ref{fig:ratio_11}) with a maximum difference of $\sim 10$\% in
the intensity ratios. This is not really surprising since the
recoupling and statistical rate coefficients for the hyperfine
transitions $1_0^+ F \to 0_0^- F'=1$ are identical and equal to the
pure rotational rate coefficients (by construction since the
rotational $0_0^-$ level has a single hyperfine sub-level
$F=1$). Thus, only the rate coefficients corresponding to hyperfine
transitions within the rotational level $1_0^+$, the so-called
quasi-elastic rate coefficients, are different in the two
datasets: they are set to zero in the statistical set while they are
comparable to the pure rotational rate coefficient in the recoupling
set. We can conclude that the quasi-elastic rate coefficients have a
significant although moderate impact in the radiative transfer
calculation.

Finally, we wish to stress that the above results are only
illustrative: more sophisticated radiative transfer calculations are
clearly needed to explore a much larger parameter space and to include
line overlap as well as systematic motion effects. Preliminary
calculations with \texttt{MOLPOP} show, for instance, that for kinetic
temperatures larger than 10~K and column densities above
$10^{15}$~cm$^{-3}$, line overlap effects can increase the intensity
ratio $F=0\to 1/F=2\to 2$ in the $1_1^--1_1^+$ inversion transition by
more than 50\%.  The combination of the present recoupling rate
coefficients with one-dimensional radiative transfer codes able to
handle hyperfine overlaps, such as \texttt{ALICO} \citep{magalhaes18},
will be particularly valuable to revisit the NH$_3$ hyperfine
anomalies and to more firmly establish the role of non-LTE effects.

\section{Conclusion}\label{sec_concl}
We have presented a full set of rate coefficient for the hyperfine excitation of \textit{ortho}- and \textit{para}-NH$_3$ in collisions with \textit{para}-H$_2$ for all levels of NH$_3$ up to $j=4$ and for temperatures up to 100 K. The calculations were performed by means of the accurate recoupling technique, which relies on quantum-mechanical close-coupling scattering calculations. In general, the rate coefficients are weakly dependent on the temperature, but depend strongly on all quantum numbers of NH$_3$. We found no clear propensity rules, and a comparison with statistical rate coefficients (which are independent of the initial hyperfine level) showed important deviations from this approximation. In addition, we also found large differences with previously published results on NH$_3$-He collisions. 

Using a simple non-LTE model, we illustrated the difference between the accurate recoupling rate coefficients and the approximate statistical rate coefficients on the inversion transition in ground state \textit{para}-NH$_3$ ($1_1^-F \rightarrow 1_1^+F'$) at 23.68 GHz and on the rotational transition $1_0^+F \rightarrow 0_0^-F'$ in \textit{ortho}-NH$_3$. These simple calculations demonstrate that the two sets of rate coefficients lead to significant differences. Moreover, non-LTE effects are predicted for column density in excess of $\sim 10^{12}$~cm$^{-2}$. 
An important extension of the present work will be to perform radiative transfer calculations using a model that can treat line overlaps which play a key role in NH$_3$ hyperfine anomalies. This would allow the investigation of non-LTE effects at higher temperatures, thus involving higher-lying rotational states of ammonia. 

As a reminder, the present calculations only considered the hyperfine splitting due to the N nucleus but did not take into account the (much smaller) further hyperfine magnetic splitting due to the three protons, which is sometimes resolved in astronomical spectra \citep{Rydbeck1977}. We suggest that the corresponding rate coefficients can be obtained from those presented here using the statistical approximation.

Finally, it should also be noted that the recoupling calculations presented in the present work are accurate but computationally expensive. At the other end, statistical rate coefficients can be immediately obtained provided the rate coefficients for pure rotational excitation are available. An intermediate approach is the IOS method, which has been used with some success for the hyperfine excitation of linear molecules by H$_2$ \citep{Goicoechea2022}. Extending the IOS method and recoupling theory to asymmetric polyatomic molecules would be beneficial in order to generate sets of hyperfine-resolved rate coefficients for other molecules such as deuterated ammonia (NH$_2$D, NHD$_2$) or H$_2$CO.

\section*{Acknowledgements}
Brian Svoboda is acknowledged for useful discussions.
J.L. acknowledges support from KU Leuven through Grant No. 19-00313.
Moreover, we acknowledge financial support from the European Research Council (Consolidator Grant COLLEXISM, Grant Agree- ment No. 811363). F.L. acknowledges financial support from the Institut Universitaire de France and the Programme National ``Physique et Chimie du Milieu Interstellaire" (PCMI) of CNRS/INSU with INC/INP cofunded by CEA and CNES.

\section*{DATA AVAILABILITY}
The data underlying this article are available upon request from the authors. The full set of rate coefficients will be made available through the EMAA database (https://emaa.osug.fr and https://dx.doi.org/10.17178/EMAA).


\bsp	
\label{lastpage}
\end{document}